\documentclass[a4paper,10pt]{article}
\usepackage[utf8x]{inputenc}

\usepackage[dvips]{geometry}
\usepackage{amsthm,amsmath,amssymb}
\usepackage{algorithmic}
\usepackage{algorithm}
\usepackage{amsmath}
\usepackage{amssymb}

\title{An expected-case sub-cubic solution to the all-pairs shortest path problem in $\mathbb R$}
\author{Julian~J.~McAuley\thanks{The authors are with the Statistical Machine Learning Program at NICTA, and the Research School of Information Sciences and Engineering, Australian National University. Queries should be addressed to \texttt{julian.mcauley@nicta.com.au}.} and Tib\'erio~S.~Caetano}

\newtheorem{theorem}{Theorem}
\newtheorem{lemma}[theorem]{Lemma}

\usepackage{cite}
\usepackage{graphicx}
\usepackage{url}

\newcommand{\eq}[1]{(eq.~\ref{#1})}

\begin{document}

\maketitle

\begin{abstract}
It has been shown by Alon et al.~that the so-called `all-pairs shortest-path' problem can be solved in $O((MV)^{2.688}\log^3(V))$ for graphs with $V$ vertices, with integer distances bounded by $M$. We solve the more general problem for graphs in $\mathbb R$ (assuming no negative cycles), with expected-case running time $O(V^{2.5}\log(V))$. While our result appears to violate the $\Omega(V^3)$ requirement of ``Funny Matrix Multiplication'' (due to Kerr), we find that it has a sub-cubic \emph{expected time} solution subject to reasonable conditions on the data distribution. The expected time solution arises when certain sub-problems are uncorrelated, though we can do better/worse than the expected-case under positive/negative correlation (respectively). Whether we observe positive/negative correlation depends on the statistics of the graph in question.
In practice, our algorithm is significantly faster than Floyd-Warshall, even for dense graphs.
\end{abstract}

\section{Problem Definition}

The all-pairs shortest path problem \cite{Dijkstra59} consists of solving
\begin{equation}
 d(v, v') = \min_{p \in \mathcal P_{v,v'}} f(p)
\label{eq:problem}
\end{equation}
for all vertices $v,v'\in\mathcal V$, where $\mathcal P_{v,v'}$ is the space of all paths connecting $v$ to $v'$ in $\mathcal V$, and $f(p)$ is the path length, i.e., $f(p) = \sum_{i=1}^{|p|-1} e(p_i, p_{i+1})$ where $e(p_i, p_j)$ is the weight of the edge connecting $p_i$ to $p_j$, or $\infty$ if no such edge exists.

A simple divide-and-conquer solution to \eq{eq:problem} can be obtained by defining $d(u, v, k)$ to be the shortest path between $u$ and $v$ containing at most $k$ edges. This solution exploits the fact that
\begin{equation}
 d(u,v,k) = \left\lbrace \begin{array}{ll}
                          e(u,v) & \text{if $k = 1$}\\
                          \min_x\left( d(u,x,k/2) + d(x,v,k/2) \right) & \text{otherwise}
                         \end{array}
\right.
\end{equation}
This allows us to solve the all-pairs shortest path problem via Algorithm \ref{alg:apsp}, which we requires $\Theta(V^3\log(V))$ time (this is by no means the optimal solution, though it is this version to which our improvements apply).

Algorithm \ref{alg:apsp}, Line \ref{line:min} requires that we solve a problem of the form
\begin{equation}
 \Phi(a,b) = \min_{x} \underbrace{\Psi_1(a,x)}_{\mathbf{v}_a} + \underbrace{\Psi_2(b,x)}_{\mathbf{v}_b}.
\label{eq:max3}
\end{equation}
Although this appears to be a \emph{linear-time} operation (in $V$), we note that it can be reduced to $O(\sqrt{V})$ (in the expected-case) if we know the permutations that sort $\mathbf{v}_a$ and $\mathbf{v}_b$. The sorted values of $\mathbf{v}_b$ will be reused for every value of $a$, and likewise the sorted values of $\mathbf{v}_a$ will be reused for every value of $b$.

Lines \ref{line:funnystart}--\ref{line:min} of Algorithm \ref{alg:apsp} are sometimes referred to as the ``Funny Matrix Multiplication'' problem: replacing $(\min, +)$ with $(+, \times)$ yields the traditional version of matrix multiplication. Kerr \cite{Kerr70} showed that it is $\Omega(V^3)$ if only the operations $\min$ and $+$ are allowed. We find that under reasonable conditions on $\mathbf{v}_a$ and $\mathbf{v}_b$, an expected-case sub-cubic solution exists, requiring only $\min$ and $+$.

\begin{algorithm}
 \caption{All-pairs shortest-path problem}
 \label{alg:apsp}
\begin{algorithmic}[1]
 \REQUIRE a graph $\mathcal V$
 \FOR{$u\in \mathcal V$}
   \FOR{$v\in \mathcal V$}
     \STATE $d(u,v,0) := e(u,v)$
   \ENDFOR
 \ENDFOR
 \FOR{$i \in \left\lbrace 1 \ldots \lceil \log V \rceil\right\rbrace$ \COMMENT{$k = 2^i$}}
   \FOR{$u\in \mathcal V$} \label{line:funnystart}
     \FOR{$v\in \mathcal V$}
       \STATE $d(u,v,i) = min_x\left( d(u,x,i-1) + d(x,v,i-1) \right)$ \COMMENT{$\Theta(V)$} \label{line:min}
     \ENDFOR
   \ENDFOR\ \COMMENT{$\Theta(V^3)$}
 \ENDFOR\ \COMMENT{$\Theta(V^3\log(V))$}
\end{algorithmic}
\end{algorithm}

\section{Our Approach}

The following elementary lemma is the key observation required in order to solve \eq{eq:max3} efficiently:

\begin{lemma}
If the $p^\text{th}$ smallest element of $\mathbf{v}_a$ has the same index as the $q^\text{th}$ smallest element of $\mathbf{v}_b$, then we only need to search through the $p$ smallest values of $\mathbf{v}_a$, and the $q$ smallest values of $\mathbf{v}_b$; any values `behind' these cannot possibly contain the smallest solution.
\label{main_lemma}
\end{lemma}

\begin{figure*}[ht]
\footnotesize
\begin{center}
\parbox[c]{31.58pt}{\centering\begin{tabular}{c}
                                     \ \\ \includegraphics[scale=0.3]{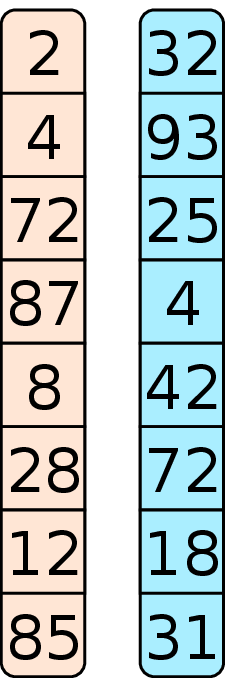}
                                    \end{tabular}
}
\parbox[c]{309.38pt}
{\centering\begin{tabular}{cccc}
$\mathit{start} = 1$ & $\mathit{start} = 2$ & $\mathit{start} = 3$ & $\mathit{start} = 4$\\
\includegraphics[scale=0.3]{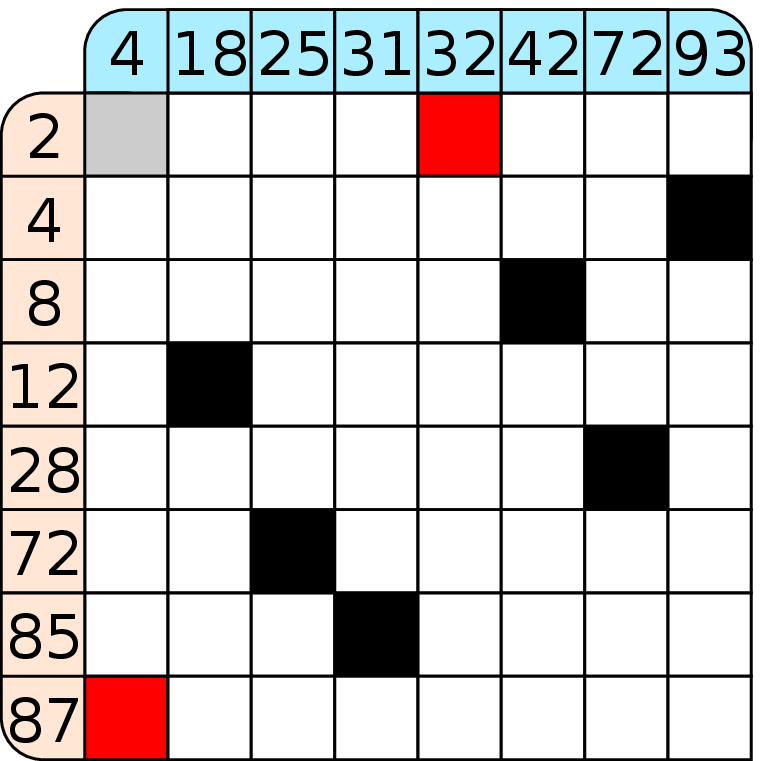} & \includegraphics[scale=0.3]{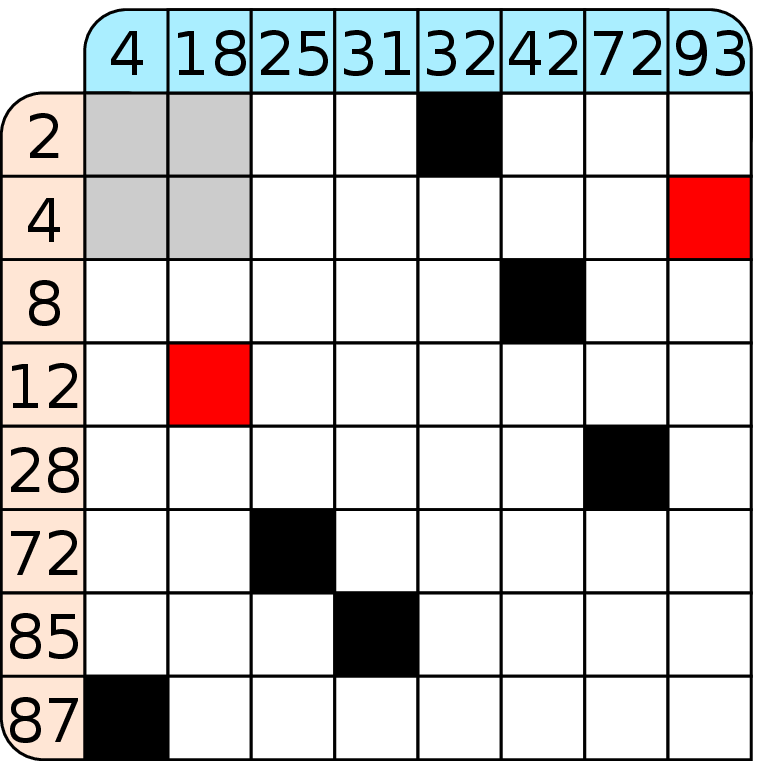} & \includegraphics[scale=0.3]{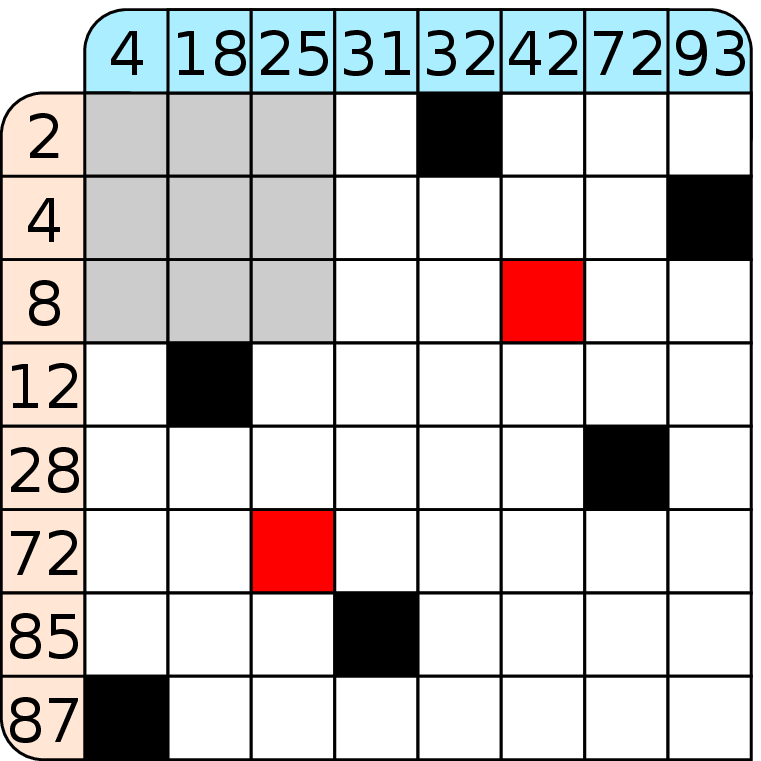} & \includegraphics[scale=0.3]{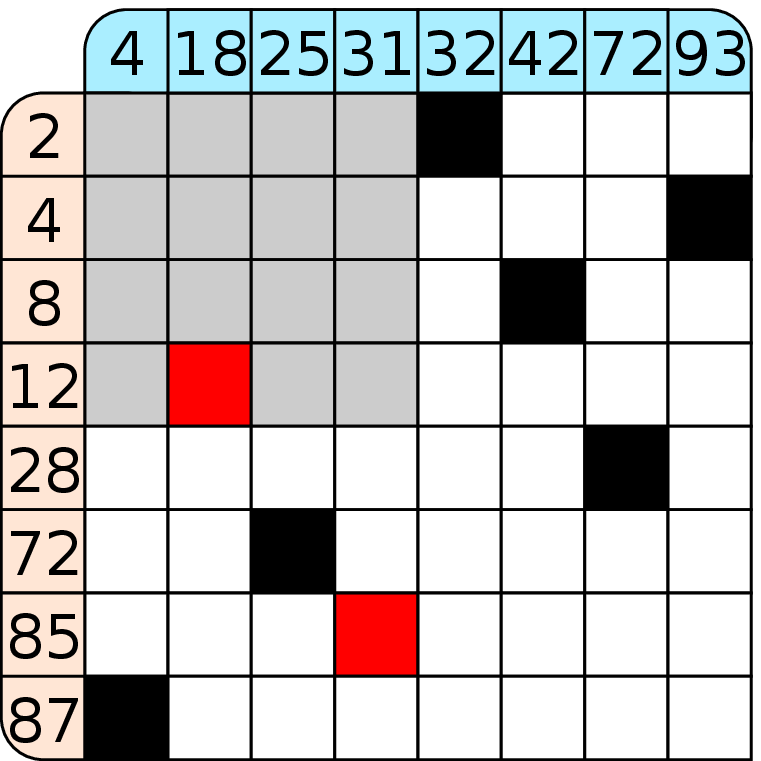}
\end{tabular}}
\end{center}
\vspace{-3mm}
 \caption{Left: The lists $\mathbf{v}_a$ and $\mathbf{v}_b$ before sorting. Right: Black squares show corresponding elements in the sorted lists ($\mathbf{v}_a[p_a[i]]$ and $\mathbf{v}_b[p_b[i]]$); red squares indicate the elements currently being read ($\mathbf{v}_a[p_a[\mathit{start}]]$ and $\mathbf{v}_b[p_b[\mathit{start}]]$). We can imagine expanding a gray box of size $\mathit{start}\times\mathit{start}$ until it contains an entry; note that the minimum is found during the first step.}
\label{fig:alg1}\end{figure*}

This observation is used to construct Algorithm \ref{alg1}. Here we iterate through the indices starting from the smallest values of $\mathbf{v}_a$ and $\mathbf{v}_b$, stopping once both indices are `behind' the minimum value found so far (which we then know is the minimum). This algorithm is demonstrated pictorially in Figure \ref{fig:alg1}.

\begin{algorithm}
  \caption{Find $i$ such that $\mathbf{v}_a[i] + \mathbf{v}_b[i]$ is minimised}
  \label{alg1}
\begin{algorithmic}[1]
\REQUIRE two vectors $\mathbf{v}_a$ and $\mathbf{v}_b$, and permutation functions $p_a$ and $p_b$ that sort them in increasing order (so that $\mathbf{v}_a[p_a[1]]$ is the smallest element in $\mathbf{v}_a$)
\STATE \textbf{Initialize:} $\mathit{start} := 1$,
$\mathit{end}_a := p_a^{-1}[p_b[1]]$, $\mathit{end}_b := p_b^{-1}[p_a[1]]$
\COMMENT{if $end_b=k$, then the smallest element in $\mathbf{v}_a$ has the same index as the $k^{\text{th}}$ smallest element in $\mathbf{v}_b$}\\
\STATE $\mathit{best} := p_a[1]$, $\mathit{min} := \mathbf{v}_a[\mathit{best}]+\mathbf{v}_b[\mathit{best}]$
\IF {$\mathbf{v}_a[p_b[1]]+\mathbf{v}_b[p_b[1]] < \mathit{min}$}
\STATE $\mathit{best} := p_b[1]$, $\mathit{min} := \mathbf{v}_a[\mathit{best}]+\mathbf{v}_b[\mathit{best}]$
\ENDIF
\WHILE{$\mathit{start} < \mathit{end}_a$
}\label{line2}
\STATE $\mathit{start} := \mathit{start} + 1$
\IF {$\mathbf{v}_a[p_a[\mathit{start}]]+\mathbf{v}_b[p_a[\mathit{start}]] < \mathit{min}$} \label{if1}
\STATE $\mathit{best} := p_a[\mathit{start}]$
\STATE $\mathit{min} := \mathbf{v}_a[\mathit{best}] + \mathbf{v}_b[\mathit{best}]$
\ENDIF
\IF {$p_b^{-1}[p_a[\mathit{start}]] < \mathit{end}_b$}
\STATE $\mathit{end}_b := p_b^{-1}[p_a[\mathit{start}]]$
\ENDIF \label{endif1}
\STATE \COMMENT{repeat Lines \ref{if1}--\ref{endif1}, interchanging $a$ and $b$} \label{line:repeat}
\ENDWHILE\ \COMMENT{this takes \emph{expected time} $O(\sqrt{V})$}
\RETURN $\mathit{best}$
\end{algorithmic}
\end{algorithm}

An upper-bound on the expected-case running time of Algorithm \ref{alg1} is given by the following theorem:
\begin{theorem}
 The \emph{expected} running time of Algorithm \ref{alg1} is $O(\sqrt{V})$.
\label{the:alg1}
\end{theorem}
The expected-case running time arises under the assumption that $\mathbf{v}_a$ and $\mathbf{v}_b$ are uncorrelated. The running time approaches $O(1)$ as $\mathbf{v}_a$ and $\mathbf{v}_b$ become increasingly correlated, and it approaches $O(V)$ as $\mathbf{v}_a$ and $\mathbf{v}_b$ become increasingly anti-correlated. Algorithm \ref{alg1} shall be analysed in detail in Section \ref{sec:analysis}.


\begin{algorithm}
 \caption{All-pairs shortest-path problem in expected-case $O(V^{2.5}\log(V))$}
 \label{alg:fast}
\begin{algorithmic}[1]
 \REQUIRE a graph $\mathcal V$
 \FOR{$u\in \mathcal V$}
   \FOR{$v\in \mathcal V$}
     \STATE $d(u,v,0) := e(u,v)$
   \ENDFOR
 \ENDFOR
 \FOR{$i \in \left\lbrace 1 \ldots \lceil \log V \rceil\right\rbrace$ \COMMENT{$k = 2^i$}}
   \FOR{$u\in \mathcal V$}
     \STATE{$p_a(u) := $ permutation that sorts $d(u,x,i-1)$}
     \STATE{$p_b(u) := $ permutation that sorts $d(x,u,i-1)$ \COMMENT{$\Theta(V\log(V))$}}
   \ENDFOR\ \COMMENT{$\Theta(V^2\log(V))$}
   \FOR{$u\in \mathcal V$}
     \FOR{$v\in \mathcal V$}
       \STATE $y := \mathit{Alg\ref{alg1}(d(u,x,i-1), d(x,v,i-1), p_a(u), p_b(v))}$ \COMMENT{$O(\sqrt{V})$}
       \STATE $d(u,v,i) := d(u,y,i-1) + d(y,v,i-1)$
     \ENDFOR
   \ENDFOR\ \COMMENT{$O(V^2\sqrt{V})$}
 \ENDFOR\ \COMMENT{$O(V^2\sqrt{V}\log(V))$}
\end{algorithmic}
\end{algorithm}

Using Algorithm \ref{alg1}, we can solve the all-pairs shortest path problem in $O(V^{2.5}\log(V))$ in the expected-case, for graphs with edge-weights in $\mathbb R$ with no negative cycles. This is shown in Algorithm \ref{alg:fast}. For dense graphs, our method has worst-case performance $\Theta(V^3\log(V))$, and best-case performance $\Theta(V^2\log^2(V))$. Our Algorithm requires $\Theta(V^2\log(V))$ memory. Also note that Algoritm \ref{alg1} can exploit sparsity in the graph structure: the algorithm may terminate as soon as it reaches entries with infinite weight -- thus if only $f(V)$ edges are viable, our algorithm has worst-case performance $O(V^2f(V)\log(V))$ (meaning that it does not surpass Johnson's Algorithm on sparse graphs \cite{johnson}).

\subsection{Comparison to Existing Approaches}

To our knowledge, the only existing sub-cubic approach is due to \cite{allpsp} (for edge weights taking small integer values); our algorithm shall not surpass this \emph{per se}, as it is not \emph{deterministic} -- it depends on the distribution of the edge weights, and it is certainly possible to adversarially generate graphs yielding worst-case performance. Our algorithm has best-case and worst-case performance of $\Theta(V^2\log^2(V))$ and $\Theta(V^3\log(V))$ respectively; thus it does not surpass Floyd-Warshall on dense graphs in the worst-case. Unlike Floyd-Warshall it is able to exploit graph sparsity, though it does not have better worst-case performance than Johnson's Algorithm. In short, our algorithm does not improve upon existing solutions in the worst-case, though under reasonable conditions, it has lower complexity than existing algorithms. We shall see in Section \ref{sec:experiments} that our algorithm is significatly faster than Floyd-Warshall in practice, making it a viable solution to real-world all-pairs shortest path problems, despite its lack of worst-case guarantees.

\section{Asymptotic Performance of Algorithm \ref{alg1}}
\label{sec:analysis}

In this section we shall determine the expected-case running time of Algorithm \ref{alg1}. Algorithm \ref{alg1} traverses $\mathbf{v}_a$ and $\mathbf{v}_b$ until it reaches the smallest value of $m$ for which there is some $j \leq m$ for which $m \geq p_b^{-1}[p_a[j]]$. If $M$ is a random variable representing this smallest value of $m$, then we wish to find $E(M)$.

\begin{figure}
\footnotesize
 \begin{center}
 \begin{tabular}{cccc}
 \includegraphics[scale=0.3]{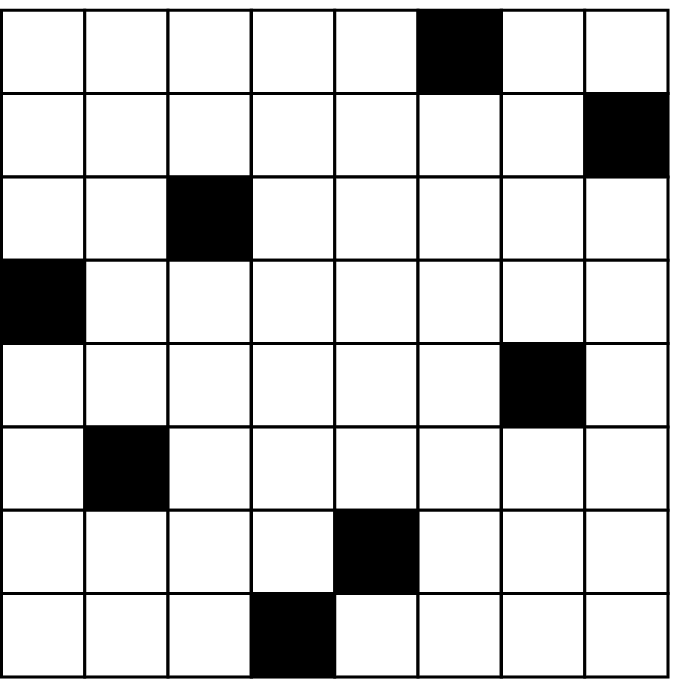} & \includegraphics[scale=0.3]{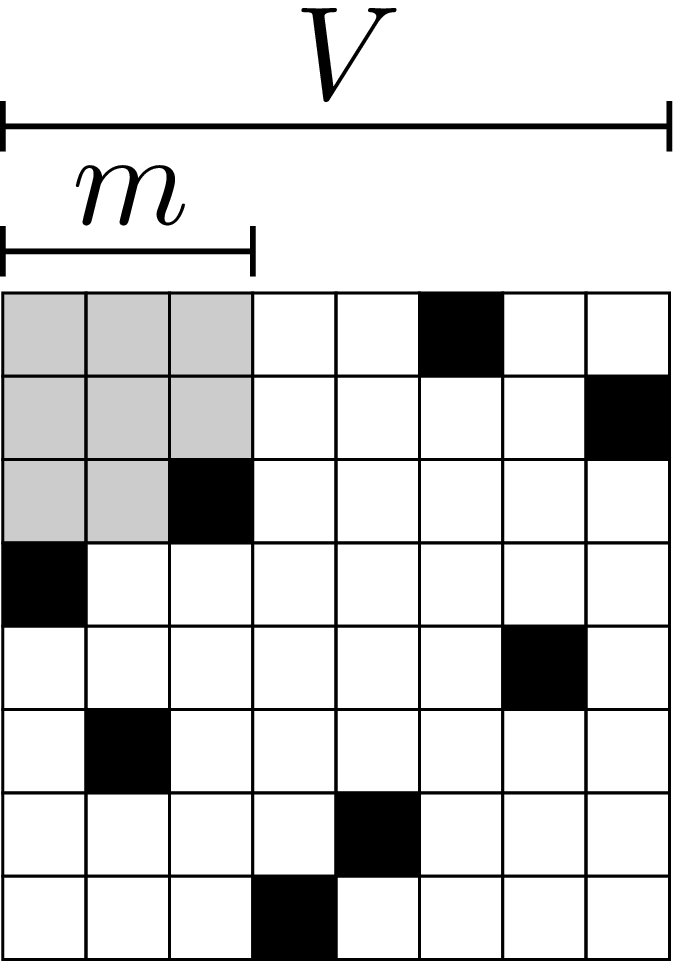} & \includegraphics[scale=0.3]{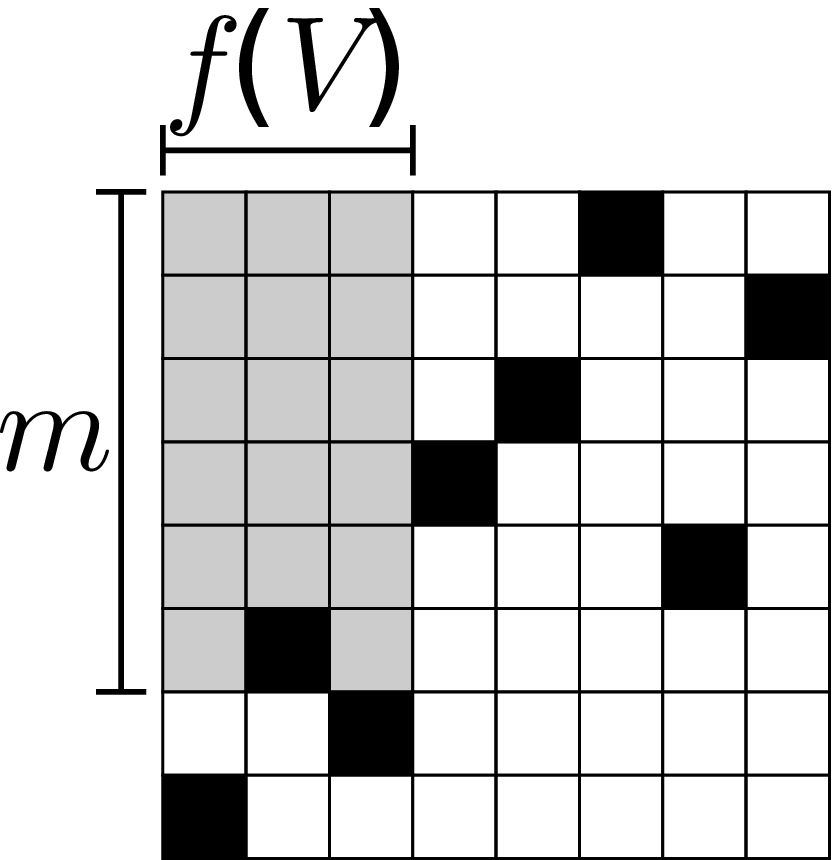}\\
(a) & (b) & (c)
 \end{tabular}
 \end{center}
\caption{(a) A permutation can be represented as an array, where there is exactly one non-zero entry in each row and column; (b) We want to find the smallest value of $m$ such that the grey box includes a non-zero entry; (c) For the sake of establishing an upper-bound, we consider a shaded region of width $f(V)$ and height $m$.}
\label{fig:perms}
\end{figure}

By representing a permutation of the digits $1$ to $V$ as shown in Figure \ref{fig:perms}, we observe that $m$ is simply the width of the smallest square (expanding from the top left) that includes an element of the permutation (i.e., it includes $i$ and $p[i]$). 


Simple analysis reveals that the probability of choosing a permutation that does not contain a value inside a square of size $m$ is
\begin{equation}
 P(M>m) = \frac{(V-m)!(V-m)!}{(V-2m)!V!}.
\label{eq:factprob}
\end{equation}
This is precisely $1 - F(m)$, where $F(m)$ is the cumulative density function of $M$. It is immediately clear that $1 \leq M \leq \lfloor V/2 \rfloor$, which defines the best and worst-case performance of Algorithm \ref{alg1}.

Using the identity $E(X) = \sum_{x=1}^\infty P(X \geq x)$, we can write down a formula for the expected value of $M$:
\begin{equation}
 E(M) = \sum_{m=0}^{\lfloor V/2 \rfloor} \frac{(V-m)!(V-m)!}{(V-2m)!V!}.
\label{eq:runtimek1}
\end{equation}
Thus the expected-case running time of our all-pairs shortest path solver (assuming uncorrelated sub-problems) is $\Theta(V^2E(M)\log(V))$. We show in the following section that $E(M)\in O(\sqrt{V})$.

\subsection{An Upper Bound on $E(M)$}

Although \eq{eq:runtimek1} precisely defines the running time of Algorithm \ref{alg1}, it is not easy to ascertain the speed improvement it achieves, as the values to which the summations converge for large $V$ are not obvious. Here, we shall try to obtain an upper-bound on their performance, which we shall assess experimentally in Section \ref{sec:experiments}. In doing so we shall prove Theorem \ref{the:alg1}.



\begin{proof}[Proof of Theorem \ref{the:alg1}]
Consider the shaded region in Figure \ref{fig:perms} (c). This region has a width of $f(V)$, and its height $m$ is chosen such that it contains precisely one non-zero value. Let $\dot{M}$ be a random variable representing the height of the grey region needed in order to include a non-zero entry. We note that
\begin{equation}
E(\dot{M}) \in O(f(V)) \rightarrow E(M) \in O(f(V));
\end{equation}
our aim is to find the smallest $f(V)$ such that $E(\dot{M}) \in O(f(V))$. The probability that none of the first $m$ samples appear in the shaded region is
\begin{equation}
 P(\dot{M} > m) = \prod_{i=0}^m \left( 1 - \frac{f(V)}{V - i}\right).
\label{eq:noreplace}
\end{equation}
Next we observe that if the entries in our $V\times V$ grid do not define a permutation, but we instead choose a \emph{random} entry in each row, then the probability (now for $\ddot{M}$) becomes
\begin{equation}
  P(\ddot{M} > m) = \left(1 - \frac{f(V)}{V}\right)^m
\label{eq:replace}
 \end{equation}
(for simplicity we allow $m$ to take arbitrarily large values). We certainly have that $P(\ddot{M}>m) \geq P(\dot{M}>m)$, meaning that $E(\ddot{M})$ is an upper bound on $E(\dot{M})$, and therefore on $E(M)$. Thus we compute the expected value
\begin{equation}
 E(\ddot{M}) = \sum_{m=0}^\infty \left(1 - \frac{f(V)}{V}\right)^m.
\end{equation}
This is just a geometric progression, which sums to ${V}/{f(V)}$. Thus we need to find $f(V)$ such that
\begin{equation}
 f(V) \in O\left(\frac{V}{f(V)}\right).
\end{equation}
Clearly $f(V) \in O(\sqrt{V})$ will do. Thus we conclude that
\begin{equation}
 E(M) \in O(\sqrt{V}).
\end{equation}
\end{proof}

We will show that this upper bound is empirically tight in the following section.

\section{Experiments}
\label{sec:experiments}

\subsection{Performance of Algorithm \ref{alg1}}

\begin{figure*}
 \begin{center}
  \includegraphics[width=0.5\textwidth]{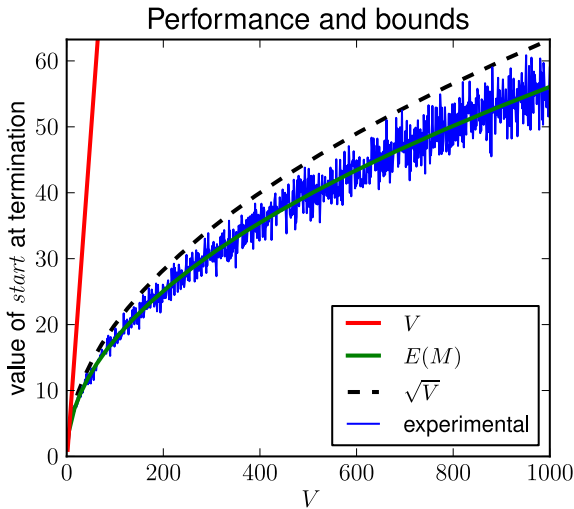}
 \end{center}
\caption{Performance of our algorithm and bounds. For $K=2$, the exact expectation is shown, which appears to precisely match the average performance (over 100 trials). The dotted lines show the upper-bound, which appears to be extremely close to the average performance, indicating that the bound is reasonably tight.}
\label{fig:exp1}
\end{figure*}

For our first experiment, we compare the performance of Algorithm \ref{alg1} to the na\"ive linear time solution. We generate $2V$ uniform samples from $[0,1)$ to obtain the lists $\mathbf{v}_a$ and $\mathbf{v}_b$. $V$ corresponds to the size of the graph in question. The performance of Algorithm \ref{alg1} is shown in Figure \ref{fig:exp1}; the value reported is simply the value of $\mathit{start}$ upon termination of the algorithm; this is compared to $V$ itself, which is the number of elements read by the na\"ive solution. The upper-bounds we obtained in the previous section are also reported, while the true expected performance (i.e., \eq{eq:runtimek1}). Visually, we find that our upper-bound is empirically very close to the true performance, suggesting that the bound is reasonably tight.

\subsection{Performance for Correlated Variables}

The expected-case running time of our algorithm was obtained under the assumption that the variables were uncorrelated, as was the case for the previous experiment. We suggested that we will obtain worse performance in the case of negatively correlated variables, and better performance in the case of positively correlated variables; we will assess these claims in this experiment.

\begin{figure*}
 \begin{center}
  \includegraphics[width=0.5\textwidth]{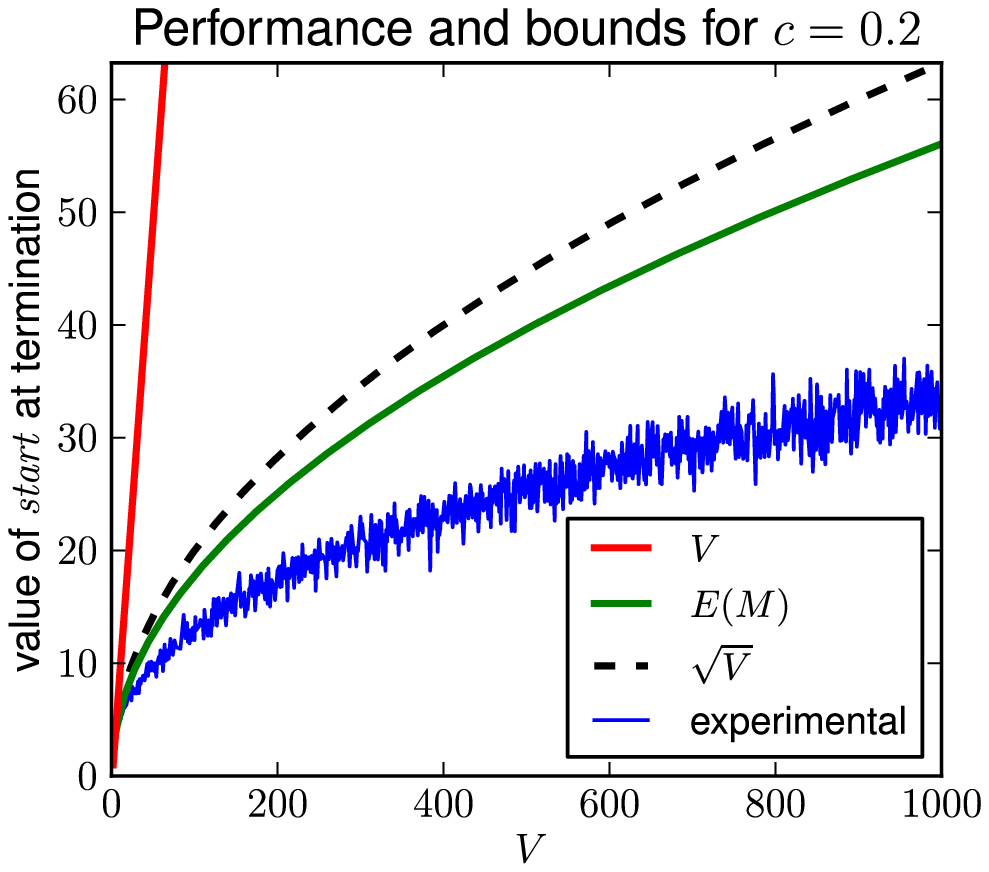}\includegraphics[width=0.5\textwidth]{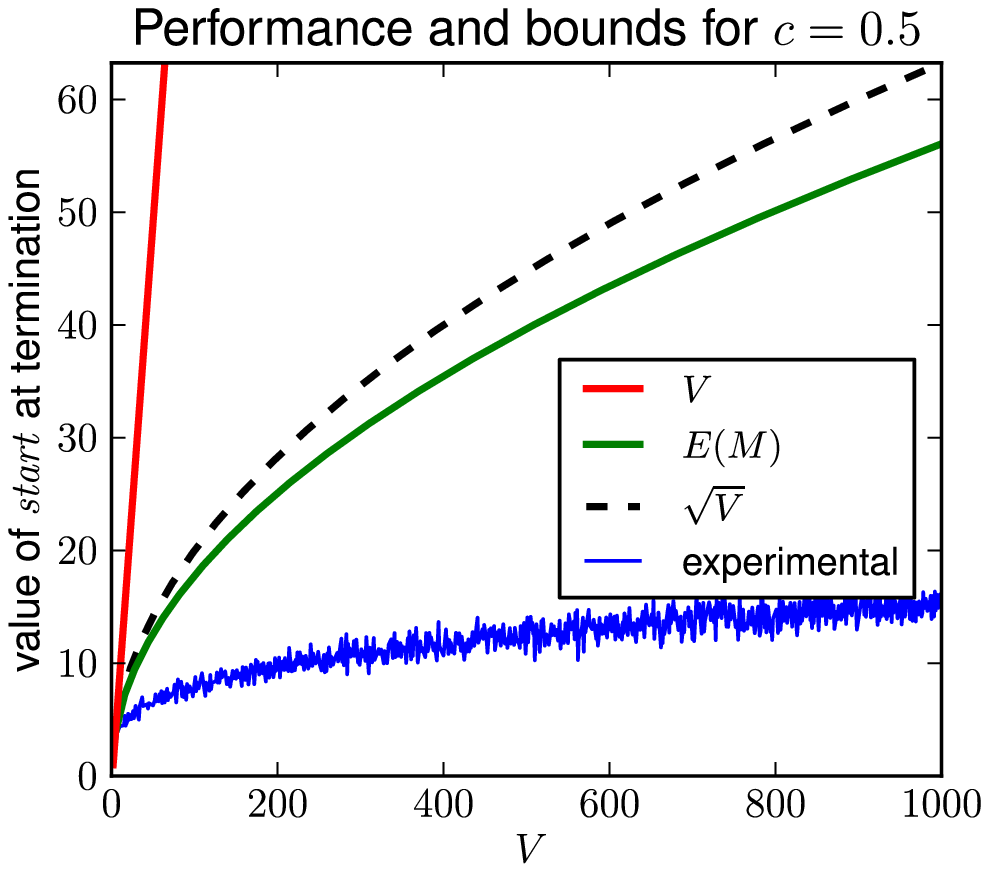}\\
  \includegraphics[width=0.5\textwidth]{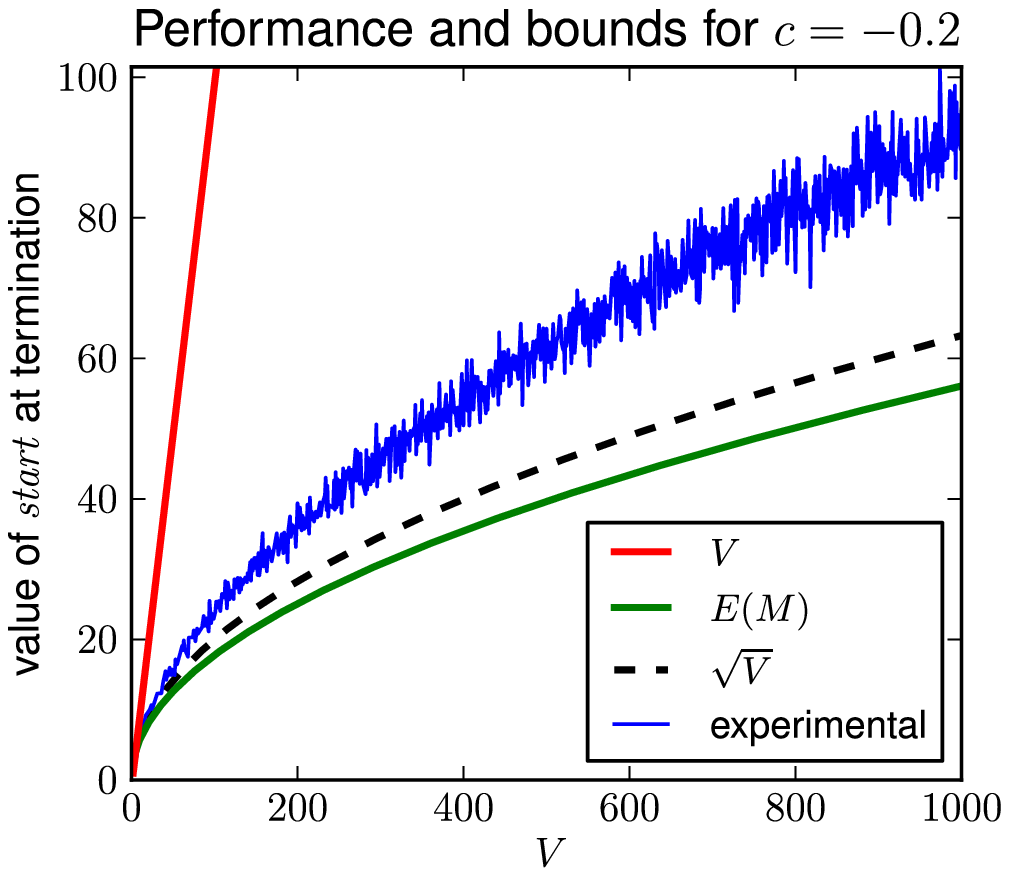}\includegraphics[width=0.5\textwidth]{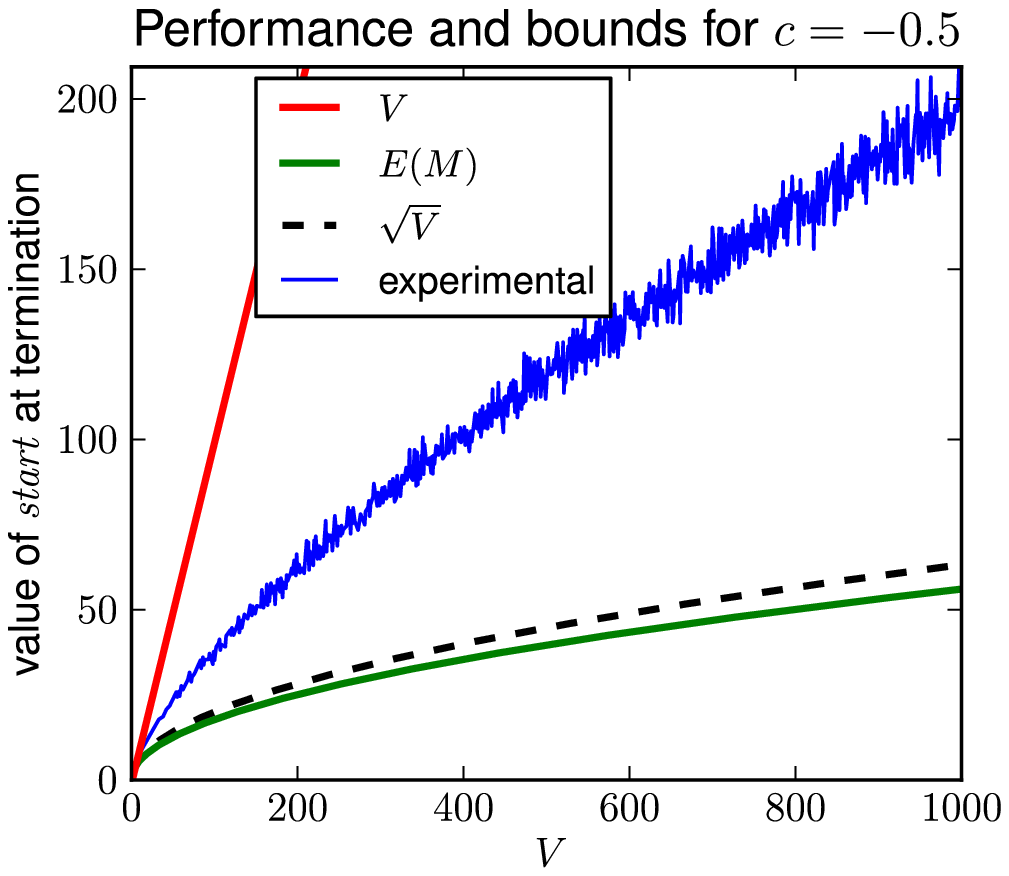}
 \end{center}
\caption{Performance of our algorithm for different correlation coefficients. The top three plots show positive correlation, the bottom three show negative correlation. Correlation coefficients of $c = 1.0$ and $c = -1.0$ capture precisely the best and worst-case performance (respectively) of our algorithm.}
\label{fig:correlated}
\end{figure*}

We report the performance for two lists (i.e., for Algorithm \ref{alg1}), whose values are sampled from a 2-dimensional Gaussian, with covariance matrix
\begin{equation}
 \Sigma = \left[ \begin{array}{cc} 1 & c \\ c & 1\end{array} \right],
\end{equation}
meaning that the two lists are correlated with correlation coefficient $c$. Performance is shown in Figure \ref{fig:correlated} for different values of $c$ ($c=0$, is not shown, as this is the case observed in the previous experiment).

In real graphs, $c$ shall be the correlation coefficient between $p(u,x,i-1)$ and $p(x,v,i-1)$ (which is free over $x$). Unless $c$ is equal to precisely $-1$ for all $u$, $v$, and $i$, we obtain a sub-cubic solution. Whether we observe positive, negative, or zero correlation will depend on the statistics of the graphs in question.

\subsection{Performance of Algorithm \ref{alg:fast}}

Finally, we compare our algorithm to the divide-and-conquer solution of Algorithm \ref{alg:apsp}, and to the popular Floyd-Warshall Algorithm \cite{floyd} on dense graphs in $\mathbb R^+$.

\begin{figure*}
 \begin{center}
  \includegraphics[width=0.5\textwidth]{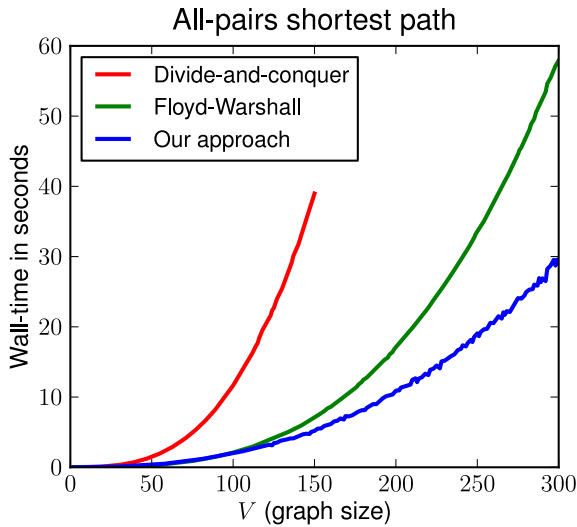}
 \end{center}
\caption{The running time of our algorithm compared to the divide-and-conquer solution of Algorithm \ref{alg:apsp}, and the Floyd-Warshall Algorithm. The average of 10 trials is shown. All algorithms were implemented in Python.}
\label{fig:compare}
\end{figure*}

We generate dense graphs of size $V$ with edge weights sampled uniformly in $[ 0, 1 )$. The performance of our algorithm, compared to Algorithm \ref{alg:apsp} and the Floyd-Warshall Algorithm is shown in Figure \ref{fig:compare}. We note that our algorithm is faster than Algorithm \ref{alg:apsp} after only $V=4$, meaning that its computational overhead is negligible. It is faster than Floyd-Warshall after $V\simeq 90$.

\subsection{Conclusion}

We have presented an expected-case subcubic solution to the problem of Funny Matrix Multiplication, resulting in an expected-case $O(V^{2.5}\log(V))$ solution to the all-pairs shortest path problem. The running time of our method depends on the distribution of edge weights for the graph in question, though we achieve performance at least as good as the expectation under reasonable conditions. Our algorithm is significantly faster than Floyd-Warshall in practice, making it a viable solution to real-world all-pairs shortest path problems.

\subsection*{Acknowledgements}

We would like to thank Pedro Felzenszwalb for alerting us to the link between inference in graphical models and the all-pairs shortest path problem. NICTA is funded by the Australian Government's \emph{Backing Australia's Ability} initiative, and the Australian Research Council's \emph{ICT Centre of Excellence} program.

\end{document}